\documentclass[twocolumn,aps,showpacs,prl,superscriptaddress]{revtex4-1}

\usepackage[T1]{fontenc}
\usepackage[utf8]{inputenc}
\usepackage{amsmath}
\usepackage{amssymb}
\usepackage{esint}
\usepackage{graphicx}   
\usepackage[normalem]{ulem}
\usepackage{color}
\usepackage{subfigure}
\usepackage{pstricks}

\begin{document}
\title{A  Monte Carlo wavefunction description of losses in a 1D Bose gas and cooling to the ground state by quantum 
feedback}
\author{M. Schemmer, A. Johnson, R. Photopoulos and I. Bouchoule}

\affiliation{Laboratoire Charles Fabry, Institut d'Optique, CNRS, Université Paris
Sud 11, 2 Avenue Augustin Fresnel, F-91127 Palaiseau Cedex, France}

\begin{abstract}
The effect of atom losses on a homogeneous one-dimensional Bose gas lying within the
quasi-condensate regime is investigated using a Monte Carlo wavefunction approach.
The evolution of the system is calculated, conditioned by the 
loss sequence, namely the times of individual losses and the position of 
the removed atoms.
We describe the gas within the linearized Bogoliubov approach.
For each mode, we find that, for a given quantum trajectory, the state of the 
system converges towards a coherent state, {\it i.e.} the ground state, displaced 
in phase space.
Provided losses are recorded with a temporal and spatially resolved detector,
we show that quantum feedback can be implemented and cooling to the ground state 
of one or several modes can be realized. 
\end{abstract}

\maketitle

In~\cite{grisins_degenerate_2016}, 
the effect of atom losses on a one-dimensional quasi-condensate was 
investigated. The authors have shown that, within a linearized approach and for a
large enough initial 
temperature, one expects the temperature of the low lying   
modes to decrease in time, in agreement with recent experimental 
results~\cite{rauer_cooling_2016}. 
The fluctuations induced by the loss process due to the discrete nature of atoms
is however responsible for a heating, limiting the temperature which can be achieved. 
More precisely, one expects 
that the temperature asymptotically converges towards 
$g\rho(t)$ where $g$ is the coupling constant and $\rho$ the linear 
atomic density~\cite{grisins_degenerate_2016,Note0}.
In particular, excitations in the phononic regime, {\it i.e.} of frequency 
much smaller than $g\rho(t)/\hbar$, never enter the quantum regime: 
their mean occupation number stays very large such that they lie in the Raighley-Jeans regime.
This heating only occurs  if one ignores the results of the losses, or, equivalently, 
if one takes the the partial trace on the state of the reservoir in which losses 
occur, ending up with the Master equation for the system's density matrix.
If on the other hand, one records the losses, more information is gained on the system 
and the analysis made in~\cite{grisins_degenerate_2016} is no longer sufficient.

 \begin{figure}[htbp]
 \centering
 \includegraphics[width=0.9\linewidth]{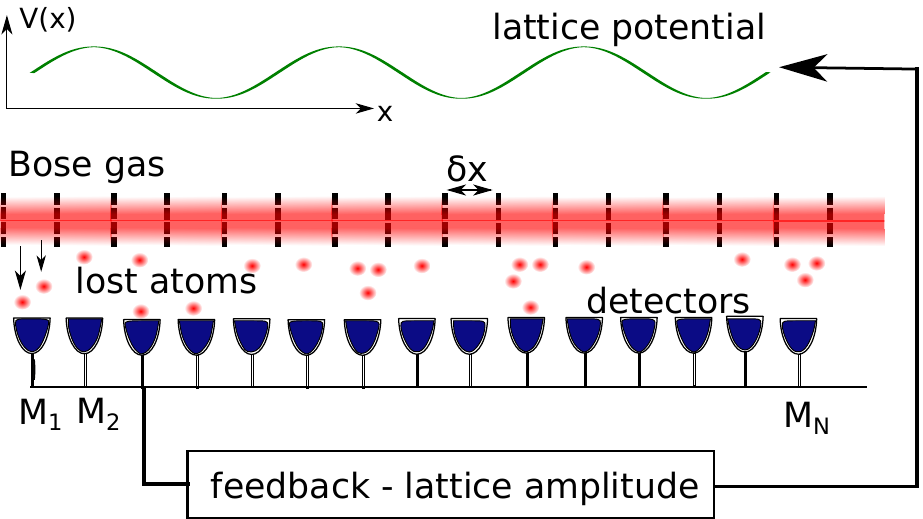}
 \caption{A one-dimensional Bose gas with atom losses and a spatially
   resolved single atom detectors system. The information on the atom
   loss can be used to create a feedback loop on the atoms via a
   lattice potential. The amplitude of the lattice potential is
   controlled by a processing unit which uses the information gained
   from the atom losses.}
 \label{fig:schema}
 \end{figure} 

In this paper, we assume the losses are monitored with a spatially
and temporally resolved 
detector 
and we  describe the evolution of the system using a Monte Carlo wavefunction
analysis. The measurement back action leads to an evolution of the system
 conditioned by the result of 
the loss process, namely  a given history of losses.
Averaging over the different possible histories, the results 
of~\cite{grisins_degenerate_2016} are recovered.
The analysis proposed in this paper however 
not only presents  an alternative picture conveying more
physical insight, but it also opens the road to the realization of
measurement based quantum feedback:
controlled dynamics, conditioned on the monitored losses,  
allows to reach lower temperatures. 
In this paper, we show that feedback on a given mode of the system 
could in principle allow to cool  this mode to the ground state. 
In particular we show that phononic excitations can be brought to the quantum regime.
 State preparation using information inferred from losses has already been used 
to prepare a well defined phase between two condensates~\cite{saba_light_2005}, 
with a  Monte Carlo wavefunction approach providing a very clear understanding 
of the mechanism~\cite{castin_relative_1997}.
Manipulation of cold atomic clouds by quantum feedback has been proposed 
in many theoretical papers using dispersive light-atom 
interaction~\cite{PhysRevA.80.013614,wade_squeezing_2015}, while feedback has been implemented 
for internal degrees of freedom~\cite{vanderbruggen_feedback_2013,vasilakis_generation_2015,kuzmich_generation_2000}.

\paragraph{Discretization of the problem.}
We consider a one-dimensional Bose gas with contact 
repulsive interactions of coupling constant $g$, such that the  
Hamiltonian writes, in second quantization, 
$$H=-\frac{\hbar^2}{2m}\!\int\! dx \psi^+\partial^2 \psi/\partial x^2 + 
\frac{g}{2}\!\int\! dx \psi^+(x)\psi^+(x)\psi(x)\psi(x).$$ 
We assume the gas is submitted to atom losses, the loss 
mechanism being a single atom process
described by a loss rate $\Gamma$. 
For instance, magnetically trapped atoms could be submitted to a radio-frequency field
that would  transfer atoms to an untrapped state, as realized experimentally in~\cite{rauer_cooling_2016}. 
Atoms could also be ionized by laser fields~\cite{kraft_spatially_2007}, 
or expelled by collision 
with fast electrons~\cite{gericke_high-resolution_2008}. We moreover assume the 
lost atoms are detected one by one 
with position-resolved detectors, as sketched in Fig.~(\ref{fig:schema}).
We note $\bar{\rho}$ the mean linear density and we 
assume the gas lies within the quasi-condensate regime such that the 
atomic density fluctuations  are small and their characteristic length scale, equal to 
the healing length $l_h=\hbar/\sqrt{mg\bar{\rho}}$, is 
much larger than the mean interparticle distance~\cite{mora_extension_2003}. 
We  discretize   space  in $N$ cells of length $\delta x$,
containing a large mean atom number $\bar{n}=\delta x \bar{\rho}$ and 
with small 
relative fluctuations. 
We  furthermore assume that $\delta x$ is large enough such that the fluctuations 
are large compared to unity\footnote{Note that these criteria can be fulfilled within the quasicondensate regime, 
for a pixel size $\delta x$ smaller than the healing length.}.
The state of the gas may be expanded (as long as one is not interested in length scales smaller than 
$\delta x$) on the Fock basis of each cell 
\begin{equation}
|\psi\rangle=\sum_{n_1,n_2,\dots ,n_N}c_{n_1,n_2,\dots ,n_N} | n_1,n_2,\dots ,n_N \rangle.
\end{equation}

\begin{figure}
\includegraphics{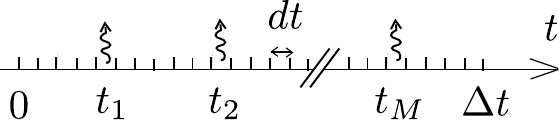}
\caption{A typical loss sequence during a time interval $\Delta t$, for a single cell. The associated 
quantum trajectory followed by the system's wavefunction is given by Eq.~\ref{eq.QMC}.}
\label{fig.QMCtraj}
\end{figure}

Time is also  discretized in intervals $\Delta t$ small compared to the time scales
involved in the longitudinal dynamics of the gas. 
This allows to consider, during $\Delta t$, the sole effect of losses first and 
then the effect of the free evolution.
We will first concentrate on the effect of losses.
Since losses do not introduce correlations between different cells,
it is relevant to consider the case of a single cell first.

\paragraph{Monte Carlo description of losses in a single cell.}
Considering a single cell, the 
initial state writes $|\psi(t)\rangle =\sum_n c_n |n\rangle$.
Let us split $\Delta t$ in elementary time steps of length
$dt$,  small enough so that the probability 
to have an atom lost  during $dt$ is small.
According to the  Monte Carlo wavefunction procedure~\cite{molmer_monte_1996},
if no atoms are detected during a time step $dt$, then the state of the 
system evolves according to the non Hermitian Hamiltonian 
$H_{\rm{eff}}=-i\hbar\Gamma a^+a/2$, which ensures the decrease of the 
 probability of  highly occupied states. If on the other hand a lost 
atom has been detected, the new state is obtained 
by the application of the jump operator $\hat{a}$, 
which annihilates an atom in the cell.
Let us now assume $M$ atoms have been lost from  the cell between time $t$ 
and time $t+\Delta t$, at times $t_1<t_2<\dots <t_M$, as sketched in Fig.~(\ref{fig.QMCtraj}).
By successively applying the  procedure described above, we
construct the quantum trajectory followed by the system and we find
\begin{equation}
\begin{array}{ll}
|\psi(t+\Delta t)\rangle=&e^{-iH_{\rm{eff}}(t+\Delta t - t_{M})/\hbar} \hat{a} e^{-iH_{\rm{eff}}(t_{M} - t_{M-1})/\hbar}  \dots\\
 &
\hat{a}e^{-iH_{\rm{eff}}(t_1-t)/\hbar}| \psi (t)\rangle.\\
\end{array}
\label{eq.QMC}
\end{equation}
With the normalization chosen here, the probability of the loss sequence is
$(dt\Gamma )^{M} \langle \psi(t+\Delta t)|\psi(t+\Delta t)\rangle/ \langle \psi(t)|\psi(t)\rangle$.
From Eq.~(\ref{eq.QMC}), we find that the Fock state coefficients 
 $c_n(t+\Delta t)=\langle n|\psi(t+\Delta t)\rangle$ write
\begin{equation}
c_n(t+\Delta t) = f_{\{t_i\}}(n+M) c_{n+M}(t), 
\end{equation}
where the function $f_{\{t_i\}}(n)$ depends on the loss sequence. 
Assuming  $\Delta t$ is small enough so that $M$ is much smaller than 
the mean atom number in the cell, itself much larger than one, 
$f_{\{t_i\}}(n)$, for a given $M$, becomes almost independent on the time sequence $\{t_i\}$
and can be approximated by 
\begin{equation}
f_{\{t_i\}}(n)\simeq f_M(n)=n^{M/2}e^{-\Gamma n t/2}.
\end{equation}
 Within this approximation, the probability of the sequence is 
$\sum_n |c_n|^2 (\Gamma n dt )^{M}e^{-\Gamma n \Delta t}$. Summing over all possible sequences 
with $M$ lost atoms, we find that the probability to have $M$ 
lost atoms is $P(M)=\sum_n |c_n|^2 (\Gamma n \Delta t )^{M}e^{-\Gamma n \Delta t}/M!$ .
For a given initial atom number $n$, 
we recover the expected Poissonian distribution.
In the limit $\bar{n}\Gamma \Delta t \gg 1$, the  typical 
number of losses is much larger than 1 and the function $f_{M}(n)$
can be approximated by the Gaussian
\begin{equation}
f_{M}(n) \simeq {\cal A}_M e^{-\frac{(n-M/(\Gamma\Delta t))^2(\Gamma \Delta t)^2 }{4M}},
\label{eq.fapp}
\end{equation}
where ${\cal A}_M$ is the normalization factor.
Using the fact that the number of lost atoms $M$, is typically equal to 
$\bar{n}\Gamma \Delta t$ and presents small relative fluctuations, 
Eq.~(\ref{eq.fapp}) further approximates to 
\begin{equation}
f_{M}(n) \simeq {\cal A}_M e^{-\frac{(n-M/(\Gamma\Delta t))^2\Gamma \Delta t }{4\bar{n}}}.
\end{equation}
The same approximations lead to a mean number of lost atoms 
$\langle M \rangle_{\rm{tr}}=\langle n\rangle \Gamma \Delta t$, with a variance 
$\langle M^2 \rangle_{\rm{tr}}-\langle M \rangle_{\rm{tr}}^2=\Gamma \Delta t \bar{n}$,
where the symbol $\rm{tr}$ indicates that averaging is done here over many different quantum trajectories.

\paragraph{Generalization to all cells and Bogoliubov decomposition.}
The results above can immediately be generalized to the case of several cells.
If $M_i$ denotes the number of lost atoms in the cell $i$, the 
probability amplitude
of the Fock state $| n_1,n_2,\dots ,n_N \rangle$ is, up to a global normalization factor, 
\begin{equation}
\begin{array}{l}
c_{n_1,n_2,\dots ,n_N}(t+\Delta t)=\\c_{n_1+M_1,n_2+ M_2,\dots ,n_N+ M_N}(t)\prod_i e^{-\frac{(n_i-M_i/(\Gamma\Delta t))^2\Gamma \Delta t }{4\bar{n}}}
\end{array}
\label{eq.modifcni}
\end{equation}
Since the atom number per cell is typically very large and present fluctuations large compared 
to unity, one can approximate discrete sums on $n_i$ by continuous integrals and treat the $n_i$ as continuous 
variables. 

Since the gas lies in the quasicondensate regime, 
its Hamiltonian is well approximated by the Bogoliubov Hamiltonian~\cite{mora_extension_2003}. 
For a homogeneous system, the Bogoliubov modes are obtained from the Fourier decomposition.
More precisely, let us introduce the Fourier quantities
\begin{equation}
\left \{
\begin{array}{l}
n_{k_i,c}=\sqrt{\frac{2}{N}}\sum_j \cos(k_i j\delta x) n_j\\
n_{k_i,s}=\sqrt{\frac{2}{N}}\sum_j \sin(k_i j\delta x) n_j\\
\end{array}
\right .
\end{equation}
 Here 
$k_i=i 2\pi/L$, where $i$ is a  integer taking values between $1$ and $(N-1)/2$. 
We introduce in the same way the operator $\hat{n}_{k_i,c}$ and $\hat{n}_{k_i,s}$.
The Bogoliubov Hamiltonian acts independently on each Fourier mode and,
for a given mode ($k,r$), where $r$ stands for $c$ or $s$ it writes, up to a constant term,  
\begin{equation}
 \hat{H}_{k,r}=A_k \hat{n}_{k,r}^2 
 +B_k \hat{\theta}_{k,r}^2,
\label{eq.HBogo}
\end{equation}
where the phase operator $\hat{\theta}_{k,r}$ is the operator conjugated to 
$\hat{n}_k$~\footnote{$[n_k,\theta_k]=i$}, 
$A_k=( g/2+\hbar^2k^2/(8m \bar{\rho}))/\delta x$  and $B_k=\hbar k^2\bar{\rho}\delta x/(2m)$ and 
the mean particle density $\bar{\rho}=\bar{n}/ \delta x$.
The frequency of the mode is $\omega_k=2\sqrt{A_k B_k}/\hbar$.

Let us now investigate the effect of losses in the Bogoliubov basis.
The state $|n_1,n_2,\dots ,n_N\rangle$ is also an eigenstate of each operator
$\hat{n}_{k_i,r}$, where $r$ stands for $c$ or $s$, with eigenvalue $n_{k_i,r}$. 
We thus use the notation $|n_1,n_2,\dots ,n_N\rangle=| \{n_{k_i,r}\} \rangle$,
where $\{n_{k_{i,r}}\}$ is a short notation for 
$n_{k_1,c},n_{k_1,s},n_{k_2,c},n_{k_2,s}\dots ,n_{k_N,c},n_{k_N,s}$.
  The state of the system then writes
\begin{equation}
|\psi\rangle=\int \prod_{i,r} dn_{k_i,r}  \tilde{c}_{\{n_{k_i,r}\}}
| \{n_{k_i,r}\} \rangle,
\end{equation}
where 
$\tilde{c}_{\{n_{k_i,r}\}} =c_{n_1,n_2,\dots ,n_N}$.
The modification of the state of the system after a 
time $\Delta t$ due to atom losses is then,
according to Eq.~(\ref{eq.modifcni}),
\begin{equation}
\tilde{c}_{\{n_{k_i,r}\}}(t+\Delta t)\!=\!\tilde{c}_{\{n_{k_i,r}+M_{k_i,r}\}}(t)\!
\prod_i e^{-\frac{(n_{k_i}-M_{k_i,r}/(\Gamma\Delta t))^2\Gamma \Delta t }{4\bar{n}}},
\label{eq.evolFourier}
\end{equation}
where $M_{k_i,c}=\sqrt{2/N}\sum_j M_j \cos(k_ix_j)$ and 
$M_{k_i,s}=\sqrt{2/N}\sum_j M_j \sin(k_ix_j)$.
We used the facts that, here, on one hand side the variances of each Gaussian in Eq.~(\ref {eq.modifcni}) 
are all equal and on the other hand side  the density profiles of Bogoliubov modes are orthogonal, namely 
the transformation between the basis ${n_i}$ and ${n_{k,r}}$ is orthogonal. 
The statistics of the different quantum trajectories gives a
Gaussian distribution for $M_{k_i,r}$ with 
$\langle M_{k_i,r}\rangle_{\rm{tr}} = \Gamma \Delta t \langle n_{k,r}\rangle$
and 
$\langle M_{k_i,r}^2\rangle_{\rm{tr}}-\langle M_{k_i,r}\rangle_{\rm{tr}}^2=\Gamma \Delta t \bar{n}$.
Eq.~(\ref{eq.evolFourier}) shows that
the losses affect each Fourier component, {\it i.e.} each Bogoliubov mode, independently. 

If the initial state is at thermal equilibrium, different Bogoliubov modes are 
 uncorrelated.
 The free evolution, under the Bogoliubov 
Hamiltonian  as well as the effect of losses,
 do not introduce correlations between modes and one can consider each 
 mode independently.  In the
following, we consider a given mode of momentum $k$ and we will omit the subscript $c$ or $s$, since 
the upcoming considerations apply for both.  

\paragraph{Evolution of a given Bogoliubov mode: Wigner representation.}
Here we  consider  a given mode, described by the two conjugate variables $n_k$ and $\theta_k$.
A convenient representation of the state of the system is its 
Wigner function $W$,
a two-dimensional real function, whose expression, as a function of the  
density matrix $D$ of the state, is
\begin{equation}
W(n_{k},\theta_k)=\frac{1}{2\pi^2}\int da db e^{i(an_k-b\theta_k)}
Tr\left ( D e^{(-ia\hat{n}_k+i b \hat{\theta}_k)}\right ).
\label{eq.wigner}
\end{equation}  
The effect of losses during $\Delta t$, in the $n_k$ representation, 
are given by Eq.~(\ref{eq.evolFourier}), 
and transform $W$ into the new Wigner function function $W'$ according to 
\begin{equation}
\begin{array}{ll}
W'(n_k,\theta_k)=&\frac{\Gamma \Delta t }{2\pi^{3/2}\bar{n}}\int d\tilde\theta 
W(n_k+M_k,\tilde \theta) \\
&e^{-\frac{\Gamma \Delta t}{2\bar{n}}(n_k-M_k/(\Gamma \Delta t))^2}
e^{-\frac{2\bar{n}}{\Gamma \Delta t}(\tilde\theta -\theta)^2}\\
\end{array}
\label{eq.transW}
\end{equation}
The multiplication by a Gaussian function along the $n_k$ axis
shifts the distribution towards $M_k/(\Gamma \Delta t)$, the value for which 
$\langle M_k \rangle_{\rm tr}$ is equal to the recorder value $M_k$.
It also
decreases the width in $n_k$, which reflects the gain of knowledge acquired on 
$n_k$ by the 
detection of the number of lost atoms.
The associated convolution along the axis $\theta_k$ 
increases the width in $\theta_k$, and ensures preservation 
of uncertainty relations.

The thermal state of the Bogoliubov Hamiltonian has a Gaussian Wigner function. 
Since the Gaussian character is preserved by  Eq.~(\ref{eq.transW}) and by the free evolution, 
the state of the system stays Gaussian. 
$W$ is then completely determined by its center $R=(\langle n_k\rangle,\langle \theta_k\rangle)$ 
and its covariance matrix
\begin{equation}
C=
\left ( 
\begin{array}{cc}
\langle n_k^2\rangle - \langle n_k\rangle^2  & \langle n_k\theta_k\rangle -\langle n_k\rangle\langle\theta_k\rangle\\
\langle n_k\theta_k\rangle -\langle n_k\rangle\langle\theta_k\rangle &
\langle \theta_k^2\rangle -\langle \theta_k\rangle^2\\
\end{array}
\right )
\end{equation}
As shown in  appendix \ref{sec.wigner}, 
to first order in $\Delta t$, the transformation in Eq.~(\ref{eq.transW}) changes $R$
and $C$ to $R'$ and $C'$ with
\begin{equation}
C'=C+\frac{\Gamma \Delta t}{\bar{n}}
\left ( 
\begin{array}{cc}
-C_{11}^2 & -C_{11}C_{12}\\
-C_{11}C_{12} & -C_{12}^2+\frac{1}{4}\\
\end{array}
\right )
\label{eq.evolcorr}
\end{equation}
and
\begin{equation}
R'=R-
\left ( 
\begin{array}{c}
\Gamma \Delta t \langle n_k\rangle\\
0\\
\end{array}
\right )
-d\xi \left ( 
\begin{array}{c}
1-C_{11}/\bar{n}\\
-C_{12}/\bar{n}\\
\end{array}\right )
\label{eq.Rprime}
\end{equation}
Here we introduced $d\xi=M_k-\Gamma \Delta t \langle n_k\rangle$.
According to the statistic of trajectories, $d\xi$ is a Gaussian variable centered on 0 and of variance 
$\langle   d\xi^2\rangle_{\rm{tr}}=\Gamma \Delta t \bar{n}$.
The above equations account for the evolution of the state under the sole
effect of atom losses. One should then implement the evolution under the 
Hamiltonian~(\ref{eq.HBogo}), which amounts to a simple rotation 
 of the Wigner function in phase space and  acts independently on $C$ and 
$R$~\footnote{The free evolution during a time $t$ amounts to a rotation 
in phase space according to the matrix  
$\mathfrak{R}(\omega_k t) = \begin{pmatrix}
\cos{\omega_k t} & \sqrt{B/A}\sin{\omega_k t} \\
-\sqrt{A/B}\sin{\omega_k t} & \cos{\omega_k t}
\end{pmatrix}$ with $\omega_k = 2 \sqrt{A B}$.  }.
Finally, one can compute the long term evolution iteratively
following the  procedure above,  knowing, at each time
interval $\Delta t$, the number of atoms lost in each cell, $M_i$, 
from which $d\xi$ is computed.  

\paragraph{Evolution of the correlation matrix.}
Eq.~(\ref{eq.evolcorr}) shows that the evolution of the correlation matrix
is the same for all possible  quantum trajectories, and general statements
can be made. Let us first consider a very slow mode such that one can ignore
the free evolution. Then $C_{12}$ stays at 0 during the evolution and   
time integration of Eq.~(\ref{eq.evolcorr}) on long times gives
\begin{equation}
C_{11}{\simeq} \bar{n}(t)/(1-e^{-\Gamma t}) , 
C_{22}{\simeq}(1-e^{-\Gamma t})/(4\bar{n}(t))
\end{equation}
where $\bar{n}(t)=\bar{n}(t=0)e^{-\Gamma t}$ is the time-dependent mean
atom-number per cell.
The system thus goes towards a state of minimal uncertainty, as 
expected, since more and more information is 
acquired on the system. 
Let us now consider the other limit of a mode of very high frequency. 
Then the free evolution of the system ensures, at any time, 
$C_{12}\simeq 0$ and the equipartition of the energy between the 2 degrees of freedom. Thus
$A_k C_{11}\simeq B_k C_{22}\simeq \langle E_{c}\rangle /2$
where  $\langle E_{c}\rangle=A_k C_{11}+B_k C_{22}$ is the contribution of the correlation matrix
to the energy. 
We then find 
\begin{equation}
\frac{d(\langle E_{c}\rangle/(\hbar\omega_k))}{dt}=\Gamma_{\rm eff} 
(
-\left ({\langle E_{c}\rangle}/(\hbar{\omega_k})\right )^2+1/4
),
\label{eq.limitComegagrand}
\end{equation}
where $\Gamma_{\rm eff}=\Gamma/\sqrt{1+4mg\bar{\rho}/(\hbar^2k^2)}$. $\Gamma_{\rm eff}$  
depends on time via 
the exponential decrease of $\bar{\rho}$ due to losses.
 At long times,  $\langle E_{c}\rangle/(\hbar\omega_k)$ goes to $1/2$, 
such that  the state of the system, as long as only the 
matrix $C$ is concerned, evolves towards the ground state. 
 If one assumes the excitation is initially in the phononic regime, however, we show in the 
appendix~\ref{SM.Ecsurgrho} 
that $\langle E_{c}\rangle$ approaches $\hbar\omega_k$ only
once the decrease of $\bar{\rho}$ has already promoted the excitation to the particle regime. Thus phononic excitations can not reach the quantum regime.
The situation is different if the decrease of $\bar{\rho}$ is compensated by the 
following time dependence of  $g$:
\begin{equation}
g(t)=g(t=0)e^{\Gamma t}.
\label{eq.gtimedep}
\end{equation}
Then $\Gamma_{\rm{eff}}$ and ${\omega_k}$ are constant and an excitation lying in the phononic regime stays
in the phononic regime during the whole loss process and, as long as the $C$ matrix is concerned, is cooled
to the ground state.
In the following, we will assume $g$ is modified according to 
Eq.~(\ref{eq.gtimedep}). 

 \begin{figure}[htbp]
 \centering
 \includegraphics[width=0.9\linewidth]{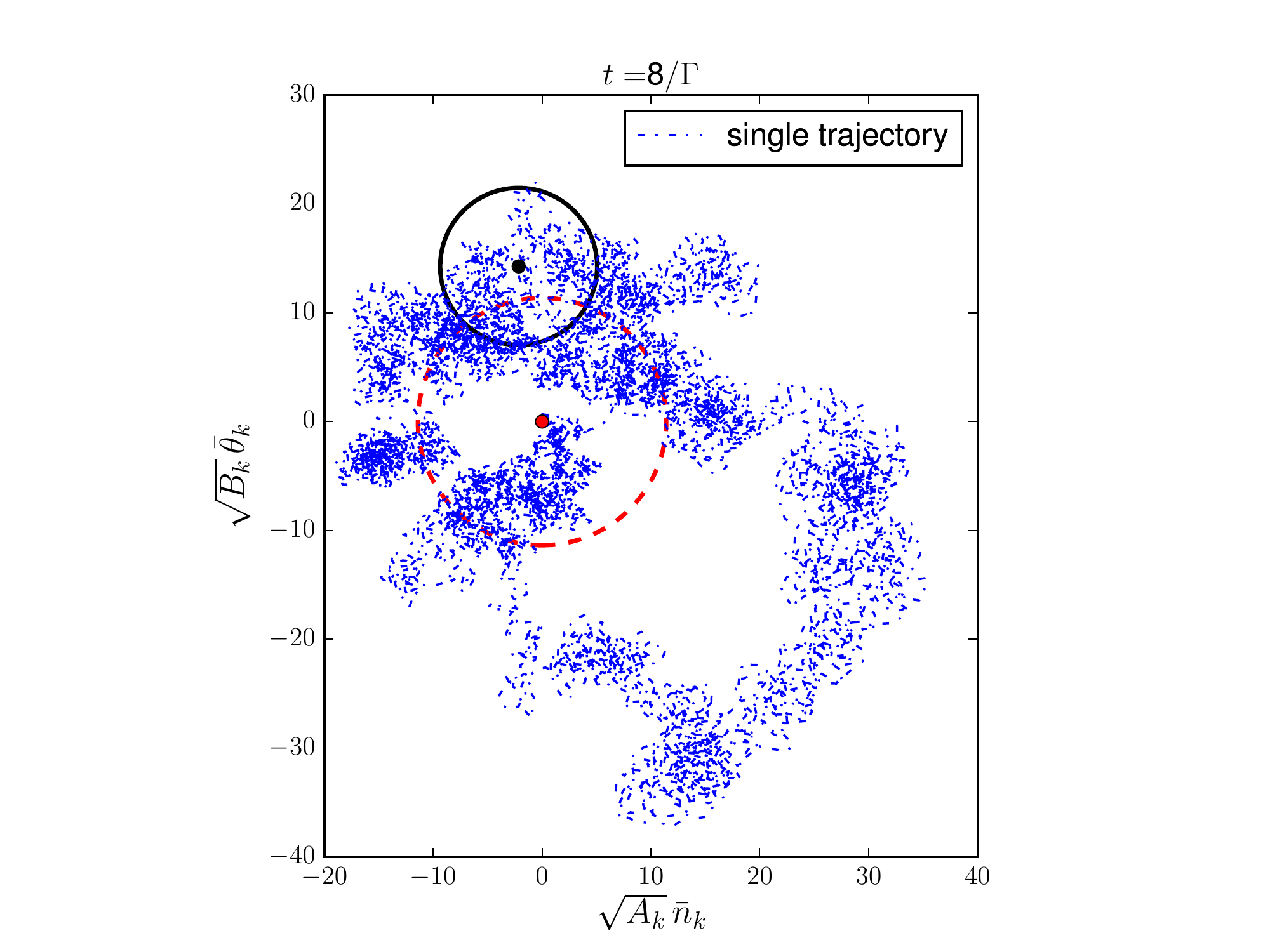}
 \caption{Evolution of the phase-space distribution for 
a single trajectory from time $t=0$ to $t=8/\Gamma$ in absence of feedback. 
The mode 
$k=0.1\sqrt{mg\bar{\rho}}/\hbar$ is considered, with  
the initial temperature $T_i= 1.1g\bar{\rho}$ and 
a loss rate $\Gamma=\omega_k/400$. Scattered blue points give the evolution 
of the center $R$.
We verified that $C_{12}$ stays small while, at any time,
$A_kC_{11}\simeq B_kC_{22} \simeq E_c/2$, as expected for such a large $\omega_k$.
 The black solid-circle, of  radius  $\sqrt{E_C}$, 
represents the final RMS width of the Wigner function.
For comparison, the red dashed-circle, of radius $\sqrt{E}$, where
$E$ is the energy obtained after averaging over 100 trajectories, gives 
the rms width of the averaged phase space distribution. 
The coordinates are given in the frame rotating according to the free evolution: namely,
the plotted quantity is 
$\bar{R}=(\bar{n}_k,\bar{\theta}_k) = \mathfrak{R}^{-1}(\omega_k t){R}$~\cite{Note3}.}
 \label{fig:ellipse}
 \end{figure}

\paragraph{Averaging over trajectories.}
If the loss events are not recorded, then only the quantities averaged over
all possible trajectories are meaningful. 
If the Wigner distribution is initially centered around 0, it will 
stay centered at 0. Let us investigate its evolution over a time $\Delta t$.
For a given quantum trajectory, {\it i.e.} a given value $d \xi$, 
the losses modifies the correlation matrix according to Eq.~(\ref{eq.evolcorr}) as 
well as the center $R$, which acquires the non zero value
$R=-d\xi (1-C_{11}/\bar{n} ; -C_{12}/\bar{n} )$. 
One then has 
\begin{equation} 
\langle n_k^2\rangle_{\rm st} (t+\Delta t)=\langle n_k^2 \rangle(t)-
\Gamma\Delta t/\bar{n} C_{11}^2+ d\xi^2 \left (1-\frac{\langle n_k^2\rangle}{\bar{n}}\right)^2
\end{equation}
and 
\begin{equation} 
\langle \theta_k^2\rangle_{\rm st} (t+\Delta t)=\langle \theta_k^2 \rangle(t)+
\Gamma\Delta t/\bar{n}(1/4-C_{12}^2) + (d\xi C_{12}/\bar{n})^2,
\end{equation}
where the subscript st specifies this holds for a single trajectory.
Averaging over all possible trajectories, we then find, 
using $\langle   d\xi^2\rangle_{\rm{tr}}=\Gamma \Delta t \bar{n}$, that losses 
modify the variances according to 
\begin{equation}
\langle n_k^2\rangle (t+\Delta t)-\langle n_k^2 \rangle(t)=
-2\Gamma \Delta t \langle n_k^2 \rangle(t)+\Gamma \Delta t \bar{n}(t).
\label{eq.evolMEnk2}
\end{equation}
and
\begin{equation}
\langle \theta_k^2\rangle (t+\Delta t)-\langle \theta_k^2 \rangle(t)=
\Gamma\Delta t/(4\bar{n}).
\label{eq.evolMEthetak2}
\end{equation}
As expected, Eq.~(\ref{eq.evolMEnk2}) and (\ref{eq.evolMEthetak2}) are equal to 
those  obtained using a master equation description of the loss 
process~\cite{grisins_degenerate_2016,tobepublished}. 
Due to the diffusive process experienced by $R$, an increased rate in both equations limits the decrease of the mode energy. 
For phonons, and assuming the loss rate is small compared to the mode frequency, 
we show in appendix~\ref{sec.Tlimit} that 
the temperature asymptotically  goes towards 
$ g\bar{\rho}(t)/2$.

This value of the asymptotic temperature $T_{\infty}$ is particular to the case of an homogeneous gas with a coupling constant evolving according to Eq.~\eqref{eq.gtimedep}. For constant $g$ the asymptotic temperature is $T_{\infty} = g \bar\rho$~\cite{grisins_degenerate_2016,tobepublished}. For a gas trapped in a harmonic potential we expect $T_{\infty}$ to scale as $g \rho_p$, where $\rho_p$ is the peak density. The proportionality factor has not been derived yet, but since the averaged density is smaller than $\rho_p$, one naively expects that $T_{\infty}$ is smaller than $g\rho_p$. Experimentally $T_{\infty}$ has not been identified, while 
temperatures as low as $0.25\,g \rho_p$ have been reported for a harmonically 
confined gas~\cite{rauer_cooling_2016}.

\paragraph{Using information retrieved from losses detection: quantum feedback.}
If the losses are recorded, such that at each time interval $\Delta t$, 
the values $M_j$ are recorded, the trajectory followed by the center of the Wigner distribution, 
$R$, can be computed exactly, and  the heating associated to the diffusion process  
seen in  equations (\ref{eq.evolMEthetak2}) and (\ref{eq.evolMEnk2})
can be compensated for.
One strategy is  to perform, during the whole time-evolution, a 
quantum feedback on the system, based on the knowledge acquired via 
the atom losses, in order to prevent the center of the Wigner distribution 
to drift away of the phase space center. 
Let us here, as an illustration, assumes one is interested in a given mode $k,c$.
The most simple back action is to submit the atomic cloud to 
a potential $V(x)=A(t)\cos(k x)$, where the computed 
amplitude $A(t)$  depends on the recorded history of the losses. 
Such a potential could be realized, for instance, 
using the dipole potential experienced by atoms in laser field.
The cosine modulation of the laser intensity can be realized using 
an optical lattice, or using a  spatial light modulator. 
The contribution to the Hamiltonian of this potential is 
$\hat{H}_{\rm{fb}}=A(t)\sqrt{N/2}\hat{n}_{k,c}$.
In order to counteract the diffusion process of $R$ due to the loss process,
one could adjust $A(t)$ such that the feedback Hamiltonian is
\begin{equation}
\hat{H}_{\rm{fb}}=-\hbar\nu \langle \theta_{k,c}\rangle \hat{n}_{k,c}
\end{equation}
where, at each time interval, $\langle \theta_{k,c}\rangle$ is computed 
by integrating the equations of 
motion  including the effect of losses, the free evolution, and the 
feedback process. This Hamiltonian acts as an active damping, 
 the damping rate $\nu$ preventing $\theta_k$ to drift far from the phase space center. 
The free evolution Hamiltonian, by coupling the two degrees of freedom, 
will ensure that neither $\theta_k$ nor $n_k$ drift away. For a large enough damping rate $\nu$,
the contribution of $R$ to the energy of the mode  is expected to 
be negligible compared to the contribution of the covariance matrix $C$ and, 
according to Eq.~(\ref{eq.limitComegagrand}), one expects to reach the ground state. 
We present numerical results illustrating such a scenario below.

 \begin{figure}[htbp]
 \centering
 \includegraphics[width=0.9\linewidth]{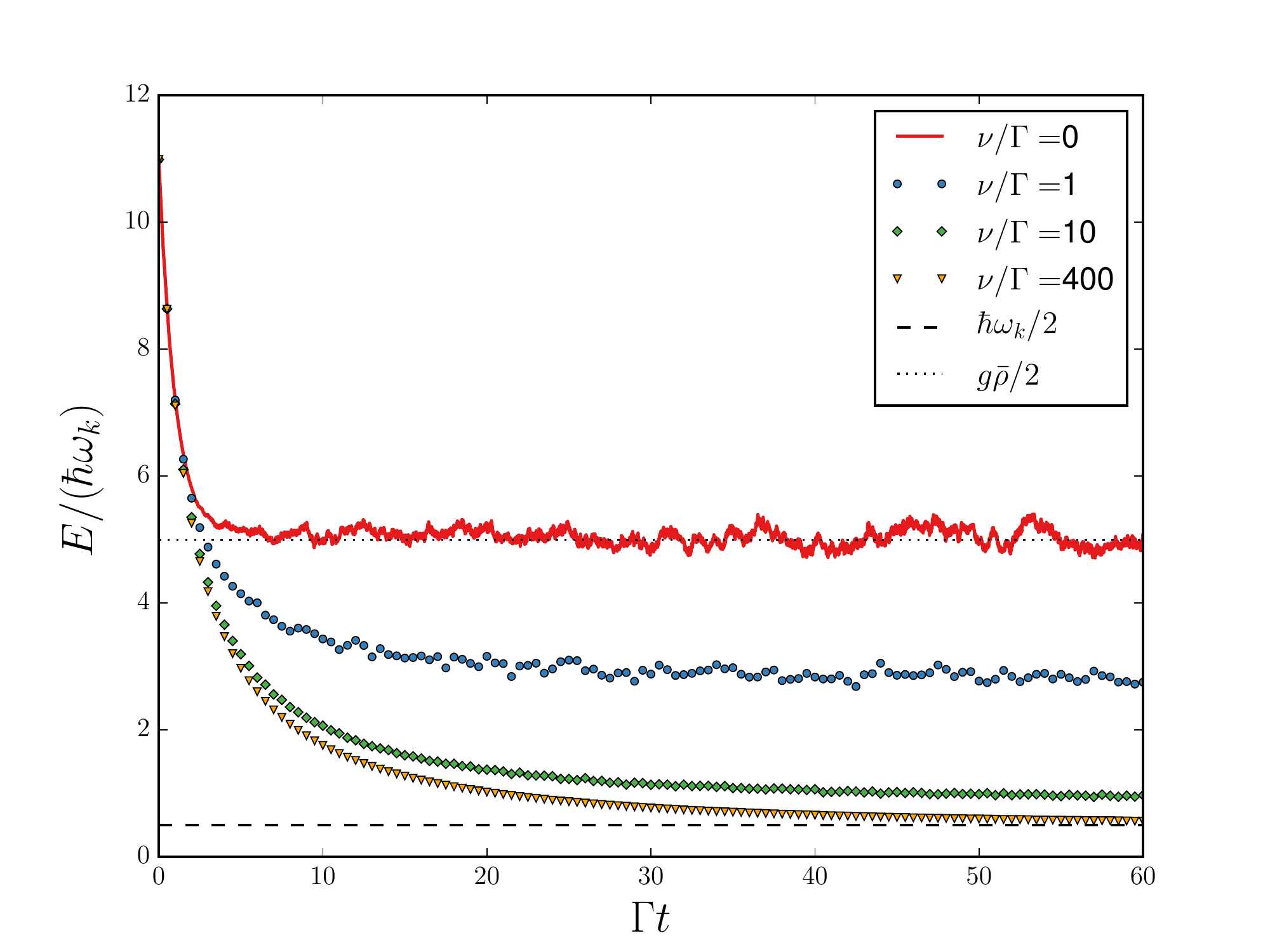}
 \caption{Simulation of the loss process for different feedback strengths $\nu$. 
Plotted is the time-evolution of the  energy in the mode $k$, averaged over 500 quantum trajectories, 
the parameters being those of Fig.~(\ref{fig:ellipse}). 
Without feedback the energy converges to $g \bar{\rho}/2$ (horizontal dotted line); 
lower energies are obtained with feedback 
and the ground state, of energy $\hbar\omega_k/2$ (horizontal dashed line) is reached for large enough $\nu$. 
 }
 \label{fig:feedback}
 \end{figure} 

Before presenting numerical results, let us identify the relevant 
quantities governing the dynamics. 
Introducing the reduced variable 
$\tilde n_k=\bar{n}^{-1/2}n_k$ and $\tilde \theta_k=\bar{n}^{1/2}\theta_k$, 
as shown in  appendix~\ref{SM.reducedvariables},
we find as expected that the cell size $\delta x$ drops out of the 
problem and, provided time is rescaled by $1/\Gamma$, the dynamics of the mode 
of wave vector $k$ is
solely governed by the dimensionless parameters $\omega_k/\Gamma$, $\hbar^2k^2/(mg\bar{\rho})$
and $\nu/\Gamma$.
The relevant measurement signal, for the time interval $\Delta t$,
is then $\tilde{M_k}=\sqrt{2/N_{at}}\int dx m(x) \cos(kx)$,
where $N_{at}$ is the total atom number and $m(x)$ is the number of lost atoms per  
unit length.  
Fig.~(\ref{fig:ellipse}), shows the phase-space evolution of a single quantum trajectory, for 
a mode lying in the phononic regime, in absence of quantum feedback. 
Fig.~(\ref{fig:feedback}) shows the time-evolution of the energy in this mode, averaged over quantum 
trajectories, both in the absence and in the presence of feedback.
In the absence of feedback,  the 
energy converges towards the expected value $g \rho/2$.
 If the feedback scheme is implemented, we observe that the 
energy in the mode reaches much smaller values. For a large feedback strength $\nu$, 
the drift of the center is almost completely prohibited and the mode is cooled 
to its ground state.

\paragraph{Discussion.}
In conclusion, we proposed a description of the effect of 
losses in a many-body system 
through a  Monte Carlo wavefunction approach, and we
showed that quantum feedback by monitoring losses
could be used to cool down selected modes of a quasi-condensate 
to vanishing temperatures. 
This work could be extended in many directions. 
In view of practical implementation, 
the sensitivity of the feedback mechanism on the exact knowledge 
of the system 
parameters should be investigated. Assuming, as is
done in this paper, the 
system parameters are known exactly, 
the larger the feedback strength, the better the cooling. 
In presence of uncertainties, a too large feedback strength will 
induce heating as it will not match the exact dynamics. 
Additionally, in most experimental situations the quasi-condensates, trapped
in a shallow longitudinal potential, are  non homogeneous.
Then, the effect of losses depends on the spatial coordinate. 
Moreover,  the linearised description should use, instead of 
the sinusoidal modes, the spatial density 
profiles of the Bogoliubov modes, which are not necessarily orthogonal.
These issues complicate the picture.  
Losses  might then induce correlations between modes~\cite{wade_squeezing_2015}. 
Another concern is the coupling between modes, which exists beyond the 
linearized approach considered here. Such coupling is present 
for instance in the Gross-Pitaevskii equation,
which is a classical field approximation of the Lieb-Liniger model. 
However, long-lived non-thermal states with different Bogoliubov modes experiencing
long life-time~\cite{langen_experimental_2015} have been reported, 
which indicates small coupling between modes and
the possibility to cool down a particular Bogoliubov mode.
Finally, note that this cooling process is not limited to 1D systems. 

\paragraph{Aknowledgment.}
The authors thanks K. M{\o}lmer for inspiring discussions. 
M. S. gratefully acknowledges support by the German Academic Scholarship Foundation.

%

\begin{thebibliography}{0}%
\makeatletter
\providecommand \@ifxundefined [1]{%
 \@ifx{#1\undefined}
}%
\providecommand \@ifnum [1]{%
 \ifnum #1\expandafter \@firstoftwo
 \else \expandafter \@secondoftwo
 \fi
}%
\providecommand \@ifx [1]{%
 \ifx #1\expandafter \@firstoftwo
 \else \expandafter \@secondoftwo
 \fi
}%
\providecommand \natexlab [1]{#1}%
\providecommand \enquote  [1]{``#1''}%
\providecommand \bibnamefont  [1]{#1}%
\providecommand \bibfnamefont [1]{#1}%
\providecommand \citenamefont [1]{#1}%
\providecommand \href@noop [0]{\@secondoftwo}%
\providecommand \href [0]{\begingroup \@sanitize@url \@href}%
\providecommand \@href[1]{\@@startlink{#1}\@@href}%
\providecommand \@@href[1]{\endgroup#1\@@endlink}%
\providecommand \@sanitize@url [0]{\catcode `\\12\catcode `\$12\catcode
  `\&12\catcode `\#12\catcode `\^12\catcode `\_12\catcode `\%12\relax}%
\providecommand \@@startlink[1]{}%
\providecommand \@@endlink[0]{}%
\providecommand \url  [0]{\begingroup\@sanitize@url \@url }%
\providecommand \@url [1]{\endgroup\@href {#1}{\urlprefix }}%
\providecommand \urlprefix  [0]{URL }%
\providecommand \Eprint [0]{\href }%
\providecommand \doibase [0]{http://dx.doi.org/}%
\providecommand \selectlanguage [0]{\@gobble}%
\providecommand \bibinfo  [0]{\@secondoftwo}%
\providecommand \bibfield  [0]{\@secondoftwo}%
\providecommand \translation [1]{[#1]}%
\providecommand \BibitemOpen [0]{}%
\providecommand \bibitemStop [0]{}%
\providecommand \bibitemNoStop [0]{.\EOS\space}%
\providecommand \EOS [0]{\spacefactor3000\relax}%
\providecommand \BibitemShut  [1]{\csname bibitem#1\endcsname}%
\let\auto@bib@innerbib\@empty
\end{thebibliography}%


\begin{thebibliography}{18}%
\makeatletter
\providecommand \@ifxundefined [1]{%
 \@ifx{#1\undefined}
}%
\providecommand \@ifnum [1]{%
 \ifnum #1\expandafter \@firstoftwo
 \else \expandafter \@secondoftwo
 \fi
}%
\providecommand \@ifx [1]{%
 \ifx #1\expandafter \@firstoftwo
 \else \expandafter \@secondoftwo
 \fi
}%
\providecommand \natexlab [1]{#1}%
\providecommand \enquote  [1]{``#1''}%
\providecommand \bibnamefont  [1]{#1}%
\providecommand \bibfnamefont [1]{#1}%
\providecommand \citenamefont [1]{#1}%
\providecommand \href@noop [0]{\@secondoftwo}%
\providecommand \href [0]{\begingroup \@sanitize@url \@href}%
\providecommand \@href[1]{\@@startlink{#1}\@@href}%
\providecommand \@@href[1]{\endgroup#1\@@endlink}%
\providecommand \@sanitize@url [0]{\catcode `\\12\catcode `\$12\catcode
  `\&12\catcode `\#12\catcode `\^12\catcode `\_12\catcode `\%12\relax}%
\providecommand \@@startlink[1]{}%
\providecommand \@@endlink[0]{}%
\providecommand \url  [0]{\begingroup\@sanitize@url \@url }%
\providecommand \@url [1]{\endgroup\@href {#1}{\urlprefix }}%
\providecommand \urlprefix  [0]{URL }%
\providecommand \Eprint [0]{\href }%
\providecommand \doibase [0]{http://dx.doi.org/}%
\providecommand \selectlanguage [0]{\@gobble}%
\providecommand \bibinfo  [0]{\@secondoftwo}%
\providecommand \bibfield  [0]{\@secondoftwo}%
\providecommand \translation [1]{[#1]}%
\providecommand \BibitemOpen [0]{}%
\providecommand \bibitemStop [0]{}%
\providecommand \bibitemNoStop [0]{.\EOS\space}%
\providecommand \EOS [0]{\spacefactor3000\relax}%
\providecommand \BibitemShut  [1]{\csname bibitem#1\endcsname}%
\let\auto@bib@innerbib\@empty
\bibitem [{\citenamefont {Grišins}\ \emph {et~al.}(2016)\citenamefont
  {Grišins}, \citenamefont {Rauer}, \citenamefont {Langen}, \citenamefont
  {Schmiedmayer},\ and\ \citenamefont {Mazets}}]{grisins_degenerate_2016}%
  \BibitemOpen
  \bibfield  {author} {\bibinfo {author} {\bibfnamefont {P.}~\bibnamefont
  {Grišins}}, \bibinfo {author} {\bibfnamefont {B.}~\bibnamefont {Rauer}},
  \bibinfo {author} {\bibfnamefont {T.}~\bibnamefont {Langen}}, \bibinfo
  {author} {\bibfnamefont {J.}~\bibnamefont {Schmiedmayer}}, \ and\ \bibinfo
  {author} {\bibfnamefont {I.~E.}\ \bibnamefont {Mazets}},\ }\href {\doibase
  10.1103/PhysRevA.93.033634} {\bibfield  {journal} {\bibinfo  {journal} {Phys.
  Rev. A}\ }\textbf {\bibinfo {volume} {93}},\ \bibinfo {pages} {033634}
  (\bibinfo {year} {2016})}\BibitemShut {NoStop}%
\bibitem [{\citenamefont {Rauer}\ \emph {et~al.}(2016)\citenamefont {Rauer},
  \citenamefont {Grišins}, \citenamefont {Mazets}, \citenamefont {Schweigler},
  \citenamefont {Rohringer}, \citenamefont {Geiger}, \citenamefont {Langen},\
  and\ \citenamefont {Schmiedmayer}}]{rauer_cooling_2016}%
  \BibitemOpen
  \bibfield  {author} {\bibinfo {author} {\bibfnamefont {B.}~\bibnamefont
  {Rauer}}, \bibinfo {author} {\bibfnamefont {P.}~\bibnamefont {Grišins}},
  \bibinfo {author} {\bibfnamefont {I.}~\bibnamefont {Mazets}}, \bibinfo
  {author} {\bibfnamefont {T.}~\bibnamefont {Schweigler}}, \bibinfo {author}
  {\bibfnamefont {W.}~\bibnamefont {Rohringer}}, \bibinfo {author}
  {\bibfnamefont {R.}~\bibnamefont {Geiger}}, \bibinfo {author} {\bibfnamefont
  {T.}~\bibnamefont {Langen}}, \ and\ \bibinfo {author} {\bibfnamefont
  {J.}~\bibnamefont {Schmiedmayer}},\ }\href {\doibase
  10.1103/PhysRevLett.116.030402} {\bibfield  {journal} {\bibinfo  {journal}
  {Phys. Rev. Lett.}\ }\textbf {\bibinfo {volume} {116}},\ \bibinfo {pages}
  {030402} (\bibinfo {year} {2016})}\BibitemShut {NoStop}%
\bibitem [{Note0()}]{Note0}%
  \BibitemOpen
  \bibinfo {note} { We express temperature in energy (effectively taking $k_B =1$).}\BibitemShut {Stop}%
\bibitem [{\citenamefont {Saba}\ \emph {et~al.}(2005)\citenamefont {Saba},
  \citenamefont {Pasquini}, \citenamefont {Sanner}, \citenamefont {Shin},
  \citenamefont {Ketterle},\ and\ \citenamefont {Pritchard}}]{saba_light_2005}%
  \BibitemOpen
  \bibfield  {author} {\bibinfo {author} {\bibfnamefont {M.}~\bibnamefont
  {Saba}}, \bibinfo {author} {\bibfnamefont {T.~A.}\ \bibnamefont {Pasquini}},
  \bibinfo {author} {\bibfnamefont {C.}~\bibnamefont {Sanner}}, \bibinfo
  {author} {\bibfnamefont {Y.}~\bibnamefont {Shin}}, \bibinfo {author}
  {\bibfnamefont {W.}~\bibnamefont {Ketterle}}, \ and\ \bibinfo {author}
  {\bibfnamefont {D.~E.}\ \bibnamefont {Pritchard}},\ }\href {\doibase
  10.1126/science.1108801} {\bibfield  {journal} {\bibinfo  {journal}
  {Science}\ }\textbf {\bibinfo {volume} {307}},\ \bibinfo {pages} {1945}
  (\bibinfo {year} {2005})}\BibitemShut {NoStop}%
\bibitem [{\citenamefont {Castin}\ and\ \citenamefont
  {Dalibard}(1997)}]{castin_relative_1997}%
  \BibitemOpen
  \bibfield  {author} {\bibinfo {author} {\bibfnamefont {Y.}~\bibnamefont
  {Castin}}\ and\ \bibinfo {author} {\bibfnamefont {J.}~\bibnamefont
  {Dalibard}},\ }\href {\doibase 10.1103/PhysRevA.55.4330} {\bibfield
  {journal} {\bibinfo  {journal} {Phys. Rev. A}\ }\textbf {\bibinfo {volume}
  {55}},\ \bibinfo {pages} {4330} (\bibinfo {year} {1997})}\BibitemShut
  {NoStop}%
\bibitem [{\citenamefont {Szigeti}\ \emph {et~al.}(2009)\citenamefont
  {Szigeti}, \citenamefont {Hush}, \citenamefont {Carvalho},\ and\
  \citenamefont {Hope}}]{PhysRevA.80.013614}%
  \BibitemOpen
  \bibfield  {author} {\bibinfo {author} {\bibfnamefont {S.~S.}\ \bibnamefont
  {Szigeti}}, \bibinfo {author} {\bibfnamefont {M.~R.}\ \bibnamefont {Hush}},
  \bibinfo {author} {\bibfnamefont {A.~R.~R.}\ \bibnamefont {Carvalho}}, \ and\
  \bibinfo {author} {\bibfnamefont {J.~J.}\ \bibnamefont {Hope}},\ }\href
  {\doibase 10.1103/PhysRevA.80.013614} {\bibfield  {journal} {\bibinfo
  {journal} {Phys. Rev. A}\ }\textbf {\bibinfo {volume} {80}},\ \bibinfo
  {pages} {013614} (\bibinfo {year} {2009})}\BibitemShut {NoStop}%
\bibitem [{\citenamefont {Wade}\ \emph {et~al.}(2015)\citenamefont {Wade},
  \citenamefont {Sherson},\ and\ \citenamefont
  {Mølmer}}]{wade_squeezing_2015}%
  \BibitemOpen
  \bibfield  {author} {\bibinfo {author} {\bibfnamefont {A.~C.}\ \bibnamefont
  {Wade}}, \bibinfo {author} {\bibfnamefont {J.~F.}\ \bibnamefont {Sherson}}, \
  and\ \bibinfo {author} {\bibfnamefont {K.}~\bibnamefont {Mølmer}},\ }\href
  {\doibase 10.1103/PhysRevLett.115.060401} {\bibfield  {journal} {\bibinfo
  {journal} {Phys. Rev. Lett.}\ }\textbf {\bibinfo {volume} {115}},\ \bibinfo
  {pages} {060401} (\bibinfo {year} {2015})}\BibitemShut {NoStop}%
\bibitem [{\citenamefont {Vanderbruggen}\ \emph {et~al.}(2013)\citenamefont
  {Vanderbruggen}, \citenamefont {Kohlhaas}, \citenamefont {Bertoldi},
  \citenamefont {Bernon}, \citenamefont {Aspect}, \citenamefont {Landragin},\
  and\ \citenamefont {Bouyer}}]{vanderbruggen_feedback_2013}%
  \BibitemOpen
  \bibfield  {author} {\bibinfo {author} {\bibfnamefont {T.}~\bibnamefont
  {Vanderbruggen}}, \bibinfo {author} {\bibfnamefont {R.}~\bibnamefont
  {Kohlhaas}}, \bibinfo {author} {\bibfnamefont {A.}~\bibnamefont {Bertoldi}},
  \bibinfo {author} {\bibfnamefont {S.}~\bibnamefont {Bernon}}, \bibinfo
  {author} {\bibfnamefont {A.}~\bibnamefont {Aspect}}, \bibinfo {author}
  {\bibfnamefont {A.}~\bibnamefont {Landragin}}, \ and\ \bibinfo {author}
  {\bibfnamefont {P.}~\bibnamefont {Bouyer}},\ }\href {\doibase
  10.1103/PhysRevLett.110.210503} {\bibfield  {journal} {\bibinfo  {journal}
  {Phys. Rev. Lett.}\ }\textbf {\bibinfo {volume} {110}},\ \bibinfo {pages}
  {210503} (\bibinfo {year} {2013})}\BibitemShut {NoStop}%
\bibitem [{\citenamefont {Vasilakis}\ \emph {et~al.}(2015)\citenamefont
  {Vasilakis}, \citenamefont {Shen}, \citenamefont {Jensen}, \citenamefont
  {Balabas}, \citenamefont {Salart}, \citenamefont {Chen},\ and\ \citenamefont
  {Polzik}}]{vasilakis_generation_2015}%
  \BibitemOpen
  \bibfield  {author} {\bibinfo {author} {\bibfnamefont {G.}~\bibnamefont
  {Vasilakis}}, \bibinfo {author} {\bibfnamefont {H.}~\bibnamefont {Shen}},
  \bibinfo {author} {\bibfnamefont {K.}~\bibnamefont {Jensen}}, \bibinfo
  {author} {\bibfnamefont {M.}~\bibnamefont {Balabas}}, \bibinfo {author}
  {\bibfnamefont {D.}~\bibnamefont {Salart}}, \bibinfo {author} {\bibfnamefont
  {B.}~\bibnamefont {Chen}}, \ and\ \bibinfo {author} {\bibfnamefont {E.~S.}\
  \bibnamefont {Polzik}},\ }\href {\doibase 10.1038/nphys3280} {\bibfield
  {journal} {\bibinfo  {journal} {Nat Phys}\ }\textbf {\bibinfo {volume}
  {11}},\ \bibinfo {pages} {389} (\bibinfo {year} {2015})}\BibitemShut
  {NoStop}%
\bibitem [{\citenamefont {Kuzmich}\ \emph {et~al.}(2000)\citenamefont
  {Kuzmich}, \citenamefont {Mandel},\ and\ \citenamefont
  {Bigelow}}]{kuzmich_generation_2000}%
  \BibitemOpen
  \bibfield  {author} {\bibinfo {author} {\bibfnamefont {A.}~\bibnamefont
  {Kuzmich}}, \bibinfo {author} {\bibfnamefont {L.}~\bibnamefont {Mandel}}, \
  and\ \bibinfo {author} {\bibfnamefont {N.~P.}\ \bibnamefont {Bigelow}},\
  }\href {\doibase 10.1103/PhysRevLett.85.1594} {\bibfield  {journal} {\bibinfo
   {journal} {Phys. Rev. Lett.}\ }\textbf {\bibinfo {volume} {85}},\ \bibinfo
  {pages} {1594} (\bibinfo {year} {2000})}\BibitemShut {NoStop}%
\bibitem [{\citenamefont {Kraft}\ \emph {et~al.}(2007)\citenamefont {Kraft},
  \citenamefont {Günther}, \citenamefont {Fortágh},\ and\ \citenamefont
  {Zimmermann}}]{kraft_spatially_2007}%
  \BibitemOpen
  \bibfield  {author} {\bibinfo {author} {\bibfnamefont {S.}~\bibnamefont
  {Kraft}}, \bibinfo {author} {\bibfnamefont {A.}~\bibnamefont {Günther}},
  \bibinfo {author} {\bibfnamefont {J.}~\bibnamefont {Fortágh}}, \ and\
  \bibinfo {author} {\bibfnamefont {C.}~\bibnamefont {Zimmermann}},\ }\href
  {\doibase 10.1103/PhysRevA.75.063605} {\bibfield  {journal} {\bibinfo
  {journal} {Phys. Rev. A}\ }\textbf {\bibinfo {volume} {75}},\ \bibinfo
  {pages} {063605} (\bibinfo {year} {2007})}\BibitemShut {NoStop}%
\bibitem [{\citenamefont {Gericke}\ \emph {et~al.}(2008)\citenamefont
  {Gericke}, \citenamefont {Würtz}, \citenamefont {Reitz}, \citenamefont
  {Langen},\ and\ \citenamefont {Ott}}]{gericke_high-resolution_2008}%
  \BibitemOpen
  \bibfield  {author} {\bibinfo {author} {\bibfnamefont {T.}~\bibnamefont
  {Gericke}}, \bibinfo {author} {\bibfnamefont {P.}~\bibnamefont {Würtz}},
  \bibinfo {author} {\bibfnamefont {D.}~\bibnamefont {Reitz}}, \bibinfo
  {author} {\bibfnamefont {T.}~\bibnamefont {Langen}}, \ and\ \bibinfo {author}
  {\bibfnamefont {H.}~\bibnamefont {Ott}},\ }\href {\doibase 10.1038/nphys1102}
  {\bibfield  {journal} {\bibinfo  {journal} {Nat Phys}\ }\textbf {\bibinfo
  {volume} {4}},\ \bibinfo {pages} {949} (\bibinfo {year} {2008})}\BibitemShut
  {NoStop}%
\bibitem [{\citenamefont {Mora}\ and\ \citenamefont
  {Castin}(2003)}]{mora_extension_2003}%
  \BibitemOpen
  \bibfield  {author} {\bibinfo {author} {\bibfnamefont {C.}~\bibnamefont
  {Mora}}\ and\ \bibinfo {author} {\bibfnamefont {Y.}~\bibnamefont {Castin}},\
  }\href {\doibase 10.1103/PhysRevA.67.053615} {\bibfield  {journal} {\bibinfo
  {journal} {Phys. Rev. A}\ }\textbf {\bibinfo {volume} {67}},\ \bibinfo
  {pages} {053615} (\bibinfo {year} {2003})}\BibitemShut {NoStop}%
\bibitem [{Note1()}]{Note1}%
  \BibitemOpen
  \bibinfo {note} {Note that these criteria can be fulfilled within the
  quasicondensate regime, for a pixel size $\delta x$ smaller than the healing
  length.}\BibitemShut {Stop}%
\bibitem [{\citenamefont {Mølmer}\ and\ \citenamefont
  {Castin}(1996)}]{molmer_monte_1996}%
  \BibitemOpen
  \bibfield  {author} {\bibinfo {author} {\bibfnamefont {K.}~\bibnamefont
  {Mølmer}}\ and\ \bibinfo {author} {\bibfnamefont {Y.}~\bibnamefont
  {Castin}},\ }\href {\doibase 10.1088/1355-5111/8/1/007} {\bibfield  {journal}
  {\bibinfo  {journal} {Quantum Semiclass. Opt.}\ }\textbf {\bibinfo {volume}
  {8}},\ \bibinfo {pages} {49} (\bibinfo {year} {1996})}\BibitemShut {NoStop}%
\bibitem [{Note2()}]{Note2}%
  \BibitemOpen
  \bibinfo {note} {$[n_k,\theta _k]=i$}\BibitemShut {NoStop}%
\bibitem [{Note3()}]{Note3}%
  \BibitemOpen
  \bibinfo {note} {The free evolution during a time $t$ amounts to a rotation
  in phase space according to the matrix $\protect \mathfrak {R}(\omega _k t) =
  \begin {pmatrix} \protect \qopname \relax o{cos}{\omega _k t} & \protect
  \sqrt {B/A}\protect \qopname \relax o{sin}{\omega _k t} \\ -\protect \sqrt
  {A/B}\protect \qopname \relax o{sin}{\omega _k t} & \protect \qopname \relax
  o{cos}{\omega _k t} \end {pmatrix}$ with $\omega _k = 2 \protect \sqrt {A
  B}$.}\BibitemShut {Stop}%
\bibitem{tobepublished}
A~Johnson, S~Szigeti, M~Schemmer, and I~Bouchoule.
\newblock {\em arXiv preprint arXiv:1703.00322}, 2017.
\bibitem [{\citenamefont {Langen}\ \emph {et~al.}(2015)\citenamefont {Langen},
  \citenamefont {Erne}, \citenamefont {Geiger}, \citenamefont {Rauer},
  \citenamefont {Schweigler}, \citenamefont {Kuhnert}, \citenamefont
  {Rohringer}, \citenamefont {Mazets}, \citenamefont {Gasenzer},\ and\
  \citenamefont {Schmiedmayer}}]{langen_experimental_2015}%
  \BibitemOpen
  \bibfield  {author} {\bibinfo {author} {\bibfnamefont {T.}~\bibnamefont
  {Langen}}, \bibinfo {author} {\bibfnamefont {S.}~\bibnamefont {Erne}},
  \bibinfo {author} {\bibfnamefont {R.}~\bibnamefont {Geiger}}, \bibinfo
  {author} {\bibfnamefont {B.}~\bibnamefont {Rauer}}, \bibinfo {author}
  {\bibfnamefont {T.}~\bibnamefont {Schweigler}}, \bibinfo {author}
  {\bibfnamefont {M.}~\bibnamefont {Kuhnert}}, \bibinfo {author} {\bibfnamefont
  {W.}~\bibnamefont {Rohringer}}, \bibinfo {author} {\bibfnamefont {I.~E.}\
  \bibnamefont {Mazets}}, \bibinfo {author} {\bibfnamefont {T.}~\bibnamefont
  {Gasenzer}}, \ and\ \bibinfo {author} {\bibfnamefont {J.}~\bibnamefont
  {Schmiedmayer}},\ }\href {\doibase 10.1126/science.1257026} {\bibfield
  {journal} {\bibinfo  {journal} {Science}\ }\textbf {\bibinfo {volume}
  {348}},\ \bibinfo {pages} {207} (\bibinfo {year} {2015})}\BibitemShut
  {NoStop}%
\end{thebibliography}

\appendix
\section{Effect of losses on the Wigner representation}
\label{sec.wigner}
Here we consider a given mode and we will omit the subscripts $k,r$
to make our notations lighter.
We also introduce $\sigma^2=\bar{n}/(\Gamma \Delta t)$ and $q_0=M_k/(\Gamma t)$.
Eq.~(\ref{eq.wigner}) writes, in representation $n$,
\begin{equation}
W(n,\theta)=\frac{1}{\pi}\int du \langle u+n|D|u-n\rangle e^{-2iu\theta}.
\end{equation}
The effect of losses, given by Eq~(\ref{eq.evolFourier})
transforms the Wigner function of the mode  to
 \begin{equation}
W'(n+M_k,\theta)=
\begin{array}[t]{l}
\frac{1}{2\pi^2\sigma^2}\int du \langle u+n|D|u-n\rangle e^{-2iu\theta}\\
 e^{-(n-q_0+u)^2/(4\sigma^2)}
e^{-(n-q_0-u)^2/(4\sigma^2)}.
\end{array}
\end{equation}
Injecting $\langle u+n|D|u-n\rangle=\int d\tilde \theta W(n,\tilde \theta)e^{i2\tilde\theta u}$, 
we then find Eq.~(\ref{eq.transW}).

Let us now consider a Gaussian state. 
Its Wigner function  writes
\begin{equation}
W(n,\theta)=\frac{1}{2\pi\sqrt{det(C)}}e^{-\frac{1}{2}\left [ (X-R)^t B (X-R) \right ]}
\label{eq.Wgaussian}
\end{equation}
where $X=\left (\begin{array}{l}n \\ \theta\end{array}\right )$, $R$ is the center of the distribution,
$C$ is the covariant matrix and $B=C^{-1}$.
The transformation in Eq.~(\ref{eq.transW}) transforms the Gaussian state 
into a new Gaussian state centered on $R'$ and of 
covariance $C'$.
The convolution on the axis $\theta$ does not change $R$ and changes  $C$ 
in $\tilde{C}$
 according to 
\begin{equation}
\tilde{C}=C+\left (\begin{array}{ll}0& 0\\0  &\frac{1}{4\sigma^2}\end{array}\right ).
\end{equation}
Let us now consider the effect  of the multiplication of $W$ by $e^{-\frac{1}{2\sigma^2}(n-q_0))^2}$,
as well as the shift along $n$ by $M_k$.
From Eq.~(\ref{eq.Wgaussian}), we find 
\begin{equation}
B'=\left ( 
\begin{array}{ll}
1/\sigma^2& 0\\
0 & 0\\
\end{array}\right )
+\tilde{B}
\label{eq.Bprime}
\end{equation}
and
\begin{equation}
\tilde{B} R+\frac{q_0}{\sigma^2} \left (\begin{array}{ll}1 \\ 0\end{array}\right )=B' 
\left ( R' + \left ( \begin{array}{l} M_k\\0\end{array}\right )
\right )
\label{eq.rprime1}
\end{equation}
where $\tilde{B}=\tilde{C}^{-1}$ and $B'=C'^{-1}$.
From Eq.~(\ref{eq.Bprime}), we obtain
\begin{equation}
C'=\left ( Id+\tilde{C}\left ( \begin{array}{ll}
1/\sigma^2& 0\\
0 & 0\\ 
\end{array}\right ) \right )^{-1} C_2
\end{equation}
 Injecting $\sigma=\sqrt{\bar{n}/(\Gamma \Delta t)}$ and expanding to first order in $\Delta t$, one 
gets
\begin{equation}
C'\simeq \left ( Id-\frac{\Gamma \Delta t}{\bar{n}} \left ( \begin{array}{ll}
C_{11}& 0\\
C_{12} & 0\\ 
\end{array}\right ) \right ) \tilde{C}.
\label{eq.Cprimeinter}
\end{equation}
Here we used the fact that $\tilde{C}_{11}=C_{11}$ and $\tilde{C}_{12}=C_{12}$.
This equation also takes the form of Eq.~(\ref{eq.evolcorr}).
Let us now consider the center of the distribution.
Multiplying the left and right hand parts of Eq.~(\ref{eq.rprime1}) by $C'$ and injecting 
(\ref{eq.Cprimeinter}), we deduce
\begin{equation}
R'=\left (Id - \frac{\Gamma \Delta t}{\bar{n}}\left ( 
\begin{array}{ll}C_{11}&0\\C_{12}&0\end{array}\right ) \right )
\left (R+\frac{M_k}{\bar{n}}\left ( \begin{array}{l}C_{11}\\C_{12}\end{array}\right )\right )
-\left ( \begin{array}{l} M_k\\0\end{array}\right )
\label{eq.rprime2}
\end{equation}
Neglecting terms beyond first order in $\Delta t$, we obtain 
\begin{equation}
R'=R+\left (\frac{M_k}{\bar{n}}-\frac{\Gamma \Delta t}{\bar{n}} \langle n_k\rangle\right )\left ( 
\begin{array}{c}
C_{11}\\
C_{12}\\
\end{array}\right )
-\left ( \begin{array}{c}
M_k\\
0\\
\end{array}\right ).
\end{equation}
Injecting $M_k=\Gamma \Delta \langle n_k\rangle +d\xi$, we recover Eq.~(\ref{eq.Rprime}).


\section{Evolution of $E_C$ for constant $g$.}\label{SM.Ecsurgrho}
We assume $\omega_k\gg \Gamma$  such that 
Eq.~(\ref{eq.limitComegagrand}) is valid. Note that the condition $\omega_k\gg \Gamma$ 
also ensures adiabatic following, namely the time evolution 
of the Hamiltonian parameters $A_k$ and $B_k$ 
preserves the ratio $E_C/(\hbar\omega_k)$, such that
 Eq.~(\ref{eq.limitComegagrand}) holds both for 
a constant $g$ and a time-varying $g$. 
Let us introduce the variable $y =  \frac{\langle E_c\rangle}{g\bar{\rho}} $ and 
rewrite Eq.~\eqref{eq.limitComegagrand} in the form 
\begin{equation}
y'(t)  =  y \Gamma \left\{ 1 -  \frac{g \rho }{2g\bar{\rho} + \hbar^2k^2/(2m)}\left(   1 +    y \right)  \right\}    + \frac{\Gamma \hbar^2k^2}{8m g \bar{\rho}}.
\end{equation}
For $y = 1 $ we see that  $y'(t) \ge 0$  and therefore $y(t)$ has to be an increasing function at $y=1$. 
It follows that for all initial conditions $y(0)\ge1$ the energy $E_C$ 
stays greater than $ g \bar{\rho}$ .
This implies in particular that, as long as an excitation stays in the 
phononic regime ({\it i.e.} its frequency stays
much smaller than $g\bar\rho/\hbar$), it stays in the high temperature regime, namely
 $E_C/(\hbar\omega_k) \gg 1$.

\section{Asymptotic temperature for non-recorded losses}
\label{sec.Tlimit}
We consider a mode $k$ (we omit the index $r$ for simplicity) and we assume averaging is done over 
trajectories.
Then evolution of the variances of $n_k$ and $\theta_k$ due to the loss process 
are given in Eq.~(\ref{eq.evolMEnk2}) and ~(\ref{eq.evolMEthetak2}). 
Let us consider the quantity $\tilde E=\langle H_k\rangle/(\hbar \omega_k)$.
We assume the loss rate is small enough so that the free evolution under the 
Hamiltonian~(\ref{eq.HBogo}) ensures  equipartition of the energy between 
the two quadratures, namely at each time 
$A_k\langle n_k^2 \rangle = B_k \langle \theta_k^2\rangle=E/2$.
Note that this is equivalent to the condition of adiabatic following.
Then the modification of $\tilde E$ under the loss process is 
\begin{equation}
\frac{1}{\Gamma}\frac{d\tilde E}{dt}=-\tilde{E}+(K+1/K)/4
\label{eq.varEtilde}
\end{equation}
where $K=4\bar{n}A_k/\omega_k=2\bar{n}\sqrt{A_k/B_k}$.
The evolution under the Hamiltonian (\ref{eq.HBogo}), provided
the adiabatic following condition is satisfied, does not modify $\tilde{E}$.
Thus Eq.~(\ref{eq.varEtilde}) gives the total time evolution of $\tilde E$, and it 
 is valid both when $A_k$ and $B_k$ depend on time 
and when they do not depend on time. 
In this paper, we consider the situation given by Eq.~(\ref{eq.gtimedep}), where the exponential
decrease of $\bar{\rho}$ is compensated by a time dependence of $g$ such that $K$ is time 
independent. Then Eq.~(\ref{eq.varEtilde}) evolves at long times to 
\begin{equation}
\tilde{E}\underset{t\rightarrow \infty}{\rightarrow}(K+1/K)/4.
\end{equation}
For phononic modes, for which $k^2\ll g\rho$, one has 
$K\simeq 2\sqrt{g\rho}/k$. Then $\tilde{E}$ goes to $\sqrt{g\rho}/(2k)$ at long times, which gives
\begin{equation}
E\underset{t\rightarrow \infty}{\rightarrow} \frac{1}{2}g\rho.
\end{equation}
This energy is very large compared to $\omega_k$. Thus the excitation lies in the 
high temperature limit and its temperature is 
$T\simeq E\simeq g\rho/2$.
Note that, in the case $g$ is constant, then $K$ depends on time and solving 
Eq.~(\ref{eq.varEtilde}) with the time-dependant value of $K$ gives that $E$ 
converges to $g\bar{\rho}=g\bar{\rho}_0 e^{-\Gamma t}$, as derived in~\cite{grisins_degenerate_2016}.

\section{Equations in reduced variables}
\label{SM.reducedvariables}
Here we derive the evolution equations for the reduced variables $\tilde n_k=n_0^{-1/2}n_k$ and $\tilde \theta_k=n_0^{1/2}\theta_k$.
We note $\tilde{R}$ and $\tilde{C}$ the associated mean 
vector and covariant matrix. 
Taking into account the exponential 
decrease  of $n_0$, Eq.~(\ref{eq.evolcorr}) and Eq.(\ref{eq.Rprime}) give
\begin{equation}
\tilde{C}'=\tilde{C}+\Gamma \Delta t
\left ( 
\begin{array}{cc}
-\tilde{C}_{11}^2+\tilde{C}_{11} & -\tilde{C}_{11}\tilde{C}_{12}\\
-\tilde{C}_{11}\tilde{C}_{12} & -\tilde{C}_{12}^2+\frac{1}{4} -\tilde{C}_{22}\\
\end{array}
\right )
\end{equation}
and
\begin{equation}
\tilde{R}'=\tilde{R}+
d\tilde{\xi}
\left ( 
\begin{array}{c}
\tilde{C}_{11}-1\\
\tilde{C}_{12}\\
\end{array}\right )
-\frac{1}{2}\left ( \begin{array}{c}
3\Gamma \Delta t \langle\tilde{n}_k\rangle \\
-\Gamma \Delta t \langle\tilde{\theta}_k\rangle \\
\end{array}\right ).
\end{equation}
Here $d\tilde{\xi}=\tilde{M}_k-\Gamma \Delta t \langle\tilde{n}_k\rangle$ where 
$\tilde{M}_k=M_k/\sqrt{n_0}$.
The statistic of trajectory implies that $\tilde{M}_k$  follows a Gaussian statistic with
 $\langle \tilde{M}_k\rangle= \Gamma \Delta t \langle\tilde{n}_k\rangle$
and $\rm{Var}\tilde{M}_k=\Gamma \delta t$.

\end{document}